\documentstyle[11pt,paspconf,psfig]{article}
\begin{document}
\title{Unified Schemes and the Two Classes of BL Lacs}
\author{P. Padovani\altaffilmark{1}\altaffilmark{,2}\altaffilmark{,3}}
\affil{Space Telescope Science Institute, Baltimore, MD, USA}
\altaffiltext{1}{Affiliated to the Astrophysics Division, Space Science 
Department, European Space Agency}
\altaffiltext{2}{On leave from Dipartimento di Fisica, II Universit\`a di 
Roma, Italy}
\altaffiltext{3}{Invited Review Talk at the {\it BL Lac Phenomenon} Meeting,
Turku, Finland, June 1998}

\begin{abstract}
I briefly summarize the main tenets of unified schemes of BL Lacs and
low-luminosity radio galaxies, discussing in particular the evolution of this
field after the Como 1988 meeting. I also examine some of the open problems
and complications of the simplest scheme. Finally, the question of the
existence of two classes of BL Lacs and our related change of perspective in
the past few years are also addressed.
\end{abstract}



\section{Unified Schemes and BL Lacs}
 
It is now well established that the appearance of Active Galactic Nuclei
(AGN) depends strongly on orientation. Classes of apparently different AGN
might actually be intrinsically similar, only viewed at different angles with
respect to the line of sight. The basic idea, based on a variety of
observations and summarized in Figure 1 of Urry \& Padovani (1995), is that
emission in the inner parts of AGN is highly anisotropic. The current paradigm
for AGN includes a central engine, surrounded by an accretion disk and by
fast-moving clouds, probably under the influence of the strong gravitational
field, emitting Doppler-broadened lines. More distant clouds emit narrower
lines. Absorbing material in some flattened configuration (usually idealized
as a torus) obscures the central parts, so that for transverse lines of sight
only the narrow-line emitting clouds are seen (narrow-lined or Type 2 AGN),
whereas the near-IR to soft-X-ray nuclear continuum and broad-lines are
visible only when viewed face-on (broad-lined or Type 1 AGN). In radio-loud
objects we have the additional presence of a relativistic jet, roughly
perpendicular to the disk, which produces strong anisotropy and amplification
of the continuum emission. 


This axisymmetric model of AGN implies widely different observational
properties (and therefore classifications) at different aspect angles. Hence
the need for ``Unified Schemes'' which look at intrinsic, isotropic
properties, to unify fundamentally identical (but apparently different)
classes of AGN. Seyfert 2 galaxies, for example, have been ``unified'' with 
Seyfert 1 galaxies (see Antonucci 1993, and references therein). 

How do BL Lacs fit into this unified picture? The first suggestion came from a
paper by Blandford \& Rees (1978) presented at the first BL Lac
conference. They proposed that many of the properties of BL Lacs could be
understood if the regions emitting continuum radiation were moving
relativistically and were viewed at relatively small angles to the line of
sight. This so-called ``relativistic beaming'' has an enormous effect on the
observed luminosities by giving rise to a very strong, angle-dependent,
amplification ($\propto \delta^p$, where $\delta$ is the Doppler factor and $p
\sim 3$), and is therefore a perfect ingredient for unified schemes.

If BL Lacs are ``beamed'' towards us, what do they look like when they are
beamed {\em away} from the observer? In other words, what is the so-called
``parent population'' of BL Lacs, i.e., that class of sources with the same
isotropic, unbeamed properties?

\subsection{From Como 1988 to Turku 1998}

By the time of the second BL Lac conference in Como, evidence was mounting
that low-luminosity Fanaroff-Riley type I (FR I) radio galaxies (Fanaroff \&
Riley 1974) were the most likely parent population of BL Lacs. This was based
on work done in the years between the first two BL Lac conferences on (rough)
estimates of the number densities of parents required (Schwartz \& Ku 1983;
Browne 1983; P\'erez-Fournon \& Biermann 1984) and studies of the extended
radio emission (Wardle et al. 1984; Antonucci \& Ulvestad 1985). This was
summarized at the meeting by Browne (1989) and additional evidence based on
host galaxy studies was given by Ulrich (1989). Not everybody at the Como
meeting thought that relativistic beaming was the correct explanation for the
properties of BL Lacs. For example, Burbidge \& Hewitt (1989) suggested that
BL Lacs had been ejected towards us from the centers of bright elliptical
galaxies, and Ostriker (1989; see also Ostriker and Vietri 1985, 1990)
proposed instead that gravitational micro-lensing by stars in a foreground
galaxy could turn an optically violently variable (OVV) quasar into a BL~Lac.

The unification picture for BL Lacs at the time of the Como meeting was
necessarily qualitative. Only by then, in fact, were the first sizeable
complete samples of BL Lacs becoming available, namely the EMSS (Maccacaro et
al. 1989), the EXOSAT (Giommi et al. 1989), and the 1 Jy (Stickel, Fried \&
K\"uhr 1989) samples. These samples (particularly the EMSS and the 1 Jy ones)
were going to play a primary role in statistical studies of BL Lacs in the
years after the Como meeting. The main change in our understanding of unified
schemes of BL Lacs after Como, in fact, was from {\em qualitative} to {\em
quantitative}. Once the first samples were available, detailed beaming models
could be tested and parameters could be inferred from the observational data.

\setcounter{footnote}{3}

In order for this scheme to be accepted, in fact, one had first to establish
that the relative numbers of objects (beamed vs. unbeamed) were correct,
taking into account the non-trivial effect of relativistic beaming (Urry \&
Shafer 1984). One approach\footnote{This is necessarily a very biased
review. Due to space limitations, I could not quote all the many papers
related to this subject.}  (taken by Meg Urry and myself) was to fit the
number counts and luminosity functions. More specifically, under the basic
assumption that AGN are randomly oriented on the sky, and assuming a radiation
pattern, one can predict the exact numbers of AGN with a given observed
luminosity relative to their intrinsic (rest-frame) luminosity. Given then the
luminosity function (LF) of the parent population, one can predict the LF of
the beamed AGN, subject to the form of the radiation pattern. More simply,
beamed objects will have higher observed powers (due to amplification) and
there will be fewer of them (because of collimation).
Fits to the radio luminosity function of 1 Jy BL Lacs, for example, based on
the luminosity function of FR I radio galaxies, indicated a distribution of
Lorentz factors $\propto \gamma^{-4}$ with $5 \la \gamma \la 30$ [$\gamma =
(1-\beta^2)^{-1/2}$, where $\beta = v/c$ is the bulk velocity in units of the
speed of light] and a critical angle separating BL Lacs from FR I radio
galaxies $\theta_{\rm c} \sim 10^{\circ}$ (Urry, Padovani \& Stickel 1991).

Another quantitative approach to the derivation of the beaming parameters
(taken by myself, Gabriele Ghisellini, Annalisa Celotti, and Laura Maraschi)
was to derive lower limits to the Doppler factor $\delta = [\gamma(1-\beta 
\cos \theta)]^{-1}$ for a large sample of radio sources. This was based on
the condition that the predicted synchrotron self Compton (SSC) flux should
not exceed the observed flux at X-ray energies. If a measurement of
superluminal velocity $\beta_{\rm app}$ is also available, one can estimate
the Lorentz factor $\gamma$ and the angle to the line of sight (with the
caveat that as the $\delta$ values are lower limits, the angles are upper
limits, while the Lorentz factors can be either depending on the relative
values of $\delta$ and $\beta_{\rm app}$). It turned out that the derived
values were consistent with those obtained from the fits to the LFs described
above (Ghisellini et al. 1993).

These studies, together with others on, for example, the distribution of
superluminal speeds in radio sources (Vermeulen \& Cohen 1994),
jet-to-counterjet ratios in radio galaxies (e.g., Giovannini et al. 1994), and
the correlation between core and total radio powers (Morganti et al. 1995),
point to a ``basic'' form of unification. This is also supported by the
evidence from the isotropic properties of BL Lacs, namely extended radio
power, host galaxy luminosity, emission line luminosity, and environment, most
of them discussed at this meeting (see also \S~1.2). In this ``zeroth-order''
unification, all BL Lacs possess jets with Lorentz factors $\langle \gamma
\rangle \simeq 3 - 7$, and are the beamed versions of FR I radio galaxies
oriented within $10^{\circ} - 20^{\circ}$ to the line of sight.

\subsection{Problems, Complications and Open Questions}
 
In the past few years, however, we have realized that this picture might be
too simplified. Indeed, various complications and open problems exist.

BL Lacs seem to avoid rich clusters (at least at low $z$), based on the
negative results obtained by Owen, Ledlow \& Keel (1996) in their search for
BL Lacs in relatively rich ($\sim 60\%$ of richness class 1 and 2) clusters at
$z < 0.09$, and on the study of the clustering environment of BL Lacs by Wurtz
et al. (1997). Interestingly, the environment of BL Lacs and their host galaxy
optical luminosity distribution (Wurtz, Stocke \& Yee 1996) seem to be more
similar to those of FR II radio galaxies. The latter comparison, however, is
misleading, as the FR I/FR II division is a function of both radio power and
absolute magnitude (e.g., Owen \& Ledlow 1994). Once this is taken into
account, BL Lacs are consistent with being beamed FR Is. (Note that the fits
to the observed luminosity functions of BL Lacs would not be acceptable if one
used only FR IIs as parents, simply because there are not enough of them.) As
regards the environment, Wurtz et al. (1996) suggest that the parent
population of BL Lacs might only include a subset of FR Is ($\sim 80\%$) which
excludes the brightest cluster galaxies in rich clusters at low $z$. This
would not substantially affect the number density results.

The vast majority of BL Lacs have extended radio powers consistent with those
of FR Is (see Urry \& Padovani 1995 and references therein). However, some
high-$z$ BL Lacs appear to have extended radio powers and radio morphologies
more consistent with FR~IIs (or rather, with what FR~IIs would look like at a
small angle with the line of sight; Kollgaard et al. 1992; Murphy et al.
1993). One problem here is that there are no FR Is at high-$z$ to compare the
BL Lacs with. In any case, this is not a serious challenge to the idea of
FR~Is being the parent population of BL~Lac objects as it affects only a
relatively few objects at the high end of the LF. A related complication
regards the arcsecond scale (VLA) radio polarization properties, as
Stanghellini et al. (1997) have shown that the polarized emission of six 1 Jy
BL Lacs, with $0.05 \la z \la 0.8$, is consistent with these sources being FR
IIs viewed end-on. This is based on the orientation of the magnetic field,
which is parallel to the jet axis. Note, however, that the polarization
properties of BL Lacs on milliarcsecond (VLBI) scales, are more typical of FR
Is than of FR IIs (Gabuzda et al. 1994).

Are FR Is the only parents of BL Lacs then? Perhaps the apparently separate
class of FR~IIs with low-excitation optical emission lines (Hine and Longair
1979; Laing et al. 1994) might be more closely associated with BL~Lac objects
than with quasars, providing a bridge between the low- and high-luminosity
unification schemes for radio loud sources. And this is obviously connected to
the relationship (if any) between BL Lacs and flat-spectrum radio quasars,
still not clear. In this respect, the study of the optical properties of the
two classes by Scarpa \& Falomo (1997) seems to suggest a continuity and even
a large overlap in emission line luminosities. 

The presence of an H$\alpha$ line in BL Lacertae (Vermeulen et al. 1995;
Corbett et al. 1996), comparable in luminosity and velocity width to that of
the Seyfert 1 galaxy NGC 4151, albeit with an equivalent width $W_{\lambda} =
5.6\pm1.4$ \AA~(so BL Lac is still a BL Lac, after all!), has called attention
to the already known fact (e.g., Miller, French \& Hawley 1978; Stickel, Fried
\& K\"uhr 1993) that some BL Lacs do have broad emission lines. This has
implications for the existence of an accretion disk in these sources and the
need for obscuration in FR Is (which lack observed broad lines).

What about micro-lensing? Is that ruled out? Apparently not, as Stocke \&
Rector (1997) interpret the $2.5 - 3 \sigma$ excess of Mg II absorbers they
find in 1 Jy BL Lacs as possible evidence that at least some of these sources
could be micro-lensed. One should consider, however, that the micro-lensing
scenario proposed by Ostriker \& Vietri was mostly a low-$z$ phenomenon
(namely, low-$z$ galaxies were turning a high-$z$ OVV into a BL Lac), while
the objects studied by Stocke \& Rector are at relatively high-$z$ ($\langle
z_{\rm abs} \rangle \sim 0.8$). I stress that there is ample evidence (e.g.,
Urry \& Padovani 1995) that micro-lensing {\em cannot} explain the properties
of the bulk of the BL Lac population.

Finally, Browne \& March\~a (1993) have suggested that recognition problems
might affect low-luminosity BL~Lacs whose light is swamped by the host galaxy,
typically a bright elliptical. A different sort of recognition problem, which
we are just starting to appreciate, is a more fundamental one, namely: How
does one define a BL Lac object? March\~a et al. (1996) have pointed out that
the ``classical'' definition of a BL Lac ($W_{\lambda} < 5$ \AA~and Ca II H
and K break $C < 0.25$) is probably too restrictive (the equivalent width
limit is somewhat arbitrary and elliptical galaxies have typically $C \sim
0.5$). They have proposed to expand upon this definition by including sources
with $C \le 0.4$ and falling in a particular area of the $W_{\lambda} - C$
plane. Such ``intermediate'' objects are now being discovered by the deeper
on-going BL Lac surveys discussed at this meeting (see also March\~a et al.
1996 and Perlman et al. 1998).

\section{Two Classes of BL Lacs?}

November 15, 1985: two closely related papers appear in the same issue of the
Astrophysical Journal, back to back. The first, {\it Optical and Radio
Properties of X-ray Selected BL Lacertae Objects}, by Stocke et al., made the
point that the properties of BL Lacs selected in the X-ray band (XBL) were
less extreme than those of BL Lacs selected in the radio band (RBL). The
suggestion was made that, within the beaming hypothesis, XBL were viewed at a
larger angle to the line of sight. The second, {\it The Radio--Optical--X-ray
Spectral Flux Distribution of Blazars}, by Ledden \& O'Dell, which dealt with
BL Lacs and highly polarized quasars (which make up the blazar class), focused
on the broad band spectral distribution.  This turned out to be bimodal, with
a majority of objects being ``X-ray normal'' and a minority of them, referred
to as ``X-ray strong blazars'', having larger (up to a factor of a thousand)
$L_{\rm x}/L_{\rm r}$ ($f_{\rm x}/f_{\rm r}$) ratios. The figures in the paper
showed that, although most ``X-ray strong'' BL Lacs were XBL, some were
actually RBL. This important point was apparently neglected for the following
ten years or so. Of the two papers, in fact, the first was the most
influential initially.

The differences between XBL and RBL can be thus summarized (see Kollgaard 1994
and Urry \& Padovani 1995 and references therein): XBL have lower optical
polarization, lower radio core-dominance, smaller optical variability, and
lower radio and optical luminosities than RBL. They also display negative
evolution (i.e., XBL were either less numerous or less luminous in the past;
Maccacaro et al. 1989; Morris et al. 1991; see also Beckmann and Giommi, Menna
\& Padovani, these proceedings), contrary to RBL, for which the evolution is
slightly positive but consistent with zero (Stickel et al. 1991). Finally, XBL
and RBL have different multifrequency spectra and therefore occupy different
regions in the $\alpha_{\rm ro} - \alpha_{\rm ox}$ plane (these are the
``usual'' radio-optical and optical-X-ray effective spectral indices).

Amongst all these differences, one similarity stood out: the average X-ray
luminosity was roughly the same for the two classes. This fact was noted by
Maraschi et al. (1986), who interpreted it as evidence that the X-ray beaming
cone was wider than the radio-optical ones. In other words, X-ray emission was
thought to be more isotropic, and therefore the observed $L_{\rm x}$ would be
not very different over a wide range of angles. Radio emission, being more
strongly beamed, would produce widely different $L_{\rm r}$, with objects seen
at smaller angles being more radio luminous. The larger the beaming cone (the
more isotropic the radiation), the easier it is to detect a source, so XBL
were predicted to be more numerous than RBL. Ghisellini \& Maraschi (1989)
gave a theoretical interpretation to this picture in terms of an accelerating
jet model, in which the X-ray emission from the most compact region would have
a smaller Lorentz factor than the more extended radio-emitting region. The
probability of detecting a source with a jet opening angle $\theta_{\rm c}$ is
$P(\theta < \theta_{\rm c}) = 1 - \cos \theta_{\rm c}$. Therefore, if
$\theta_{\rm c,r}$ and $\theta_{\rm c,x}$ are the radio and X-ray opening
angles of BL Lacs, the ratio between the number densities of the two classes
will be $N_{\rm XBL}/N_{\rm RBL} = (1 - \cos \theta_{\rm c,x})/(1 - \cos 
\theta_{\rm c,r})$. With the values for the angles derived from statistical 
studies ($\theta_{\rm c,x} \sim 30^{\circ}$ and $\theta_{\rm c,r} \sim
10^{\circ}$: see Urry \& Padovani 1995), it turns out that XBL should be about
10 times more numerous than RBL, in rough agreement with the X-ray number
counts for the two classes (Urry, Padovani \& Stickel 1991).

Our picture of BL Lac unification by 1993, summarized in Figure 7 of
Ghi\-sel\-lini et al. (1993), had then RBL as seen within $\sim 15^{\circ}$
with respect to the line of sight, XBL at larger angles, up to $\sim
30^{\circ}$, and FR Is occupying the remaining angles (this will be referred
to in the following as the ``different viewing angle'' hypothesis). As
discussed above, this tied in with number densities and a theoretical jet
model, so everybody was happy with it!

\subsection{From XBL/RBL to HBL/LBL}

In 1994, however, Paolo Giommi and I made the following point (Giommi \&
Padovani 1994): when the EMSS and 1 Jy samples were the only sizeable BL Lac
samples available, the division between XBL and RBL was clear-cut. However,
with the advent of all-sky X-ray surveys, like the Slew survey (Perlman et
al. 1996) and especially the {\it ROSAT} All-Sky Survey (RASS; Voges et
al. 1996), this was not the case anymore. Some RBL belonging to the 1 Jy
sample now were also XBL! How were they then supposed to be classified:
RBL/XBL? It was clear that a distinction based on the band of selection was
not physical and bound to collapse with the advent of deeper radio and X-ray
surveys.

We then went back to what we thought was a more fundamental way of
distinguishing BL Lacs, namely their broad-band spectra. This was similar to
the Ledden \& O'Dell approach, but ten years after their paper many more
multifrequency data were available. Giommi, Ansari \& Micol (1995) were then
able to quantify the differences between the spectra in terms of different
frequencies of the peak of the synchrotron emission, $\nu_{\rm peak}$. Namely,
it turned out that {\em most} XBL had $\nu_{\rm peak}$ in the UV/X-ray band,
while {\em most} RBL peaked at IR/optical energies. Since $\nu_{\rm peak}$ is
not easy to determine for the majority of the sources (which have only 2 -- 3
multifrequency data points), Paolo Giommi and I distinguished the two types of
BL Lacs on the basis of their X-ray-to-radio flux ratio $f_{\rm x}/f_{\rm r}$,
a parameter strongly correlated with $\nu_{\rm peak}$ and much easier to
derive. High-energy peaked BL Lacs, or HBL, had $f_{\rm x}/f_{\rm r} \ga
10^{-11}$ (where $f_{\rm x}$ is in erg cm$^{-2}$ s$^{-1}$ and $f_{\rm r}$ is
in Jy), while low-energy peaked BL Lacs, or LBL, had lower $f_{\rm x}/f_{\rm
r}$ values (Padovani \& Giommi 1995). The dividing value was based on the
observed $f_{\rm x}/f_{\rm r}$ distribution of the 1 Jy and EMSS samples,
which exhibited a gap around this value (see Fig. 2 of Padovani \& Giommi
1995). Our view, however, was that there was a continuous distribution of peak
frequencies (or, alternatively, $f_{\rm x}/f_{\rm r}$ values), and that the
dichotomy was a selection effect.

But that was only the beginning. Once we got rid of the XBL/RBL division and
turned to something more physical, problems with the ``different viewing
angle'' hypothesis started to emerge. X-ray selection will obviously favor BL
Lacs with peak emission at UV/X-ray energies, and so will find fewer which
peak at lower energies. But then the fact that LBL are rare in X-ray surveys
cannot tell us anything about the relative abundance of the two classes! In
fact, if radio rather than X-ray surveys are unbiased (because the radio
emission does not ``know'' the position of the peak of the emission) then HBL
are relatively rare, about $10\%$ in the 1~Jy/S4/S5 radio-selected samples. In
essence, we took the opposite approach from Maraschi et al.  (1986), assuming
radio selection rather than X-ray selection is unbiased, and found the
opposite result: in complete contrast to the accelerating jet picture, we
concluded that HBL made up only $10\%$ of the BL Lac population, and not
$90\%$!  Specifically, we argued that HBL outnumber LBL at a given X-ray flux,
even though they are intrinsically less numerous, because the two classes
sample different parts of the BL~Lac radio counts. As a consequence of their
higher $f_{\rm x}/f_{\rm r}$ ratios, HBL have lower radio fluxes ($\sim 10$
mJy\footnote{This obviously refers to currently known sources. Unified schemes
predict that BL Lacs should reach much fainter radio fluxes.} as compared to
$\sim 1$ Jy for LBL) and since fainter objects are more numerous than brighter
ones (the radio counts are rising), their surface density is higher. Stated
differently, X-ray surveys sample the BL~Lac radio counts at low fluxes and
mostly detect the $\sim 10\%$ of objects with high $f_{\rm x}/f_{\rm r}$
ratios. This holds down to quite faint X-ray fluxes, well below the {\it
ROSAT} deep survey limits, below which the fraction of LBL should increase
slowly and eventually dominate by a factor $\sim 10$.

Our hypothesis, which literally turned upside down our view of the two classes
of BL Lacs, was put to test. Starting from the observed properties of LBL (and
with no free parameters), we were able to explain most of the properties of
HBL, namely their X-ray number counts and LF and their $f_{\rm r}$
distribution. The bimodal distribution of BL Lacs in the $\alpha_{\rm ro} -
\alpha_{\rm ox}$ plane was also explained as an obvious result of the peak of 
the emission moving from high frequencies ($\sim 10^{17}$ Hz) for HBL to low
frequencies ($\sim 10^{12}$ Hz) for LBL.

The reaction to our idea was initially strong but (slowly) people started to
realize that, after all, we might not be completely wrong!  How do we explain
the similar average $L_{\rm x}$ for the two BL Lac classes, which after all
had been one of the main observational bases of the previous hypothesis? We
think that the large $L_{\rm x}/L_{\rm r}$ values for HBL compensate for their
small radio powers and conspire to give $L_{\rm x}$ values similar to those of
LBL. Another way to look at this is to notice that, while the radio fluxes of
the 1 Jy sample, on one side, and EMSS and Slew samples, on the other side,
span almost four orders of magnitude, the X-ray fluxes cover only about two
orders of magnitude. Coupled with the fact that the mean redshifts for the two
classes are within a factor of 2, it follows that the X-ray luminosities are
bound to be similar. Of course this will not necessarily be the case for
deeper BL Lac samples, so that is going to be an important test of this idea.

What about the accelerating jet model, which constituted the theoretical basis
for the ``different viewing angle'' hypothesis? Sambruna, Maraschi \& Urry
(1996) applied this model to the multifrequency spectra of the 1 Jy and EMSS
BL Lacs. They showed that it was impossible to explain the large difference in
peak frequencies ($4-5$ orders of magnitude) between HBL and LBL only by
changing the viewing angle. The observed sequence of BL Lac spectral
properties required instead a systematic change of intrinsic physical
parameters, such as magnetic field, jet size, and maximum electron energy.

We now have another argument which suggests that at least some HBL cannot be
seen at large angles to the line of sight: their $\gamma$-ray emission. If
a source is very compact, all gamma-rays are absorbed through photon--photon
collisions and produce electron -- positron pairs. But if the radiation is
beamed, the intrinsic photon density is much lower (by a factor $\delta^{-3} -
\delta^{-4}$) and gamma-ray photons manage to escape (e.g., Maraschi,
Ghisellini \& Celotti 1992). The latest list of EGRET blazars (Mukherjee et
al. 1997), which contains 51 sources, includes (besides 11 LBL) 3 HBL, namely
S5 0716+714, MKN 421 and PKS 2155$-$304. These objects should have relatively
high values of $\delta$ (because otherwise they would not be GeV sources:
Ghisellini 1997) and therefore should be seen at small angles, as for any
value of the Lorentz factor $\sin \theta \le 1/\delta$ (Urry \& Padovani
1995). Moreover, the only four sources detected at TeV energies are MKN 421,
MKN 501, 1ES 2344+514, and PKS 2155$-$304 (the latter being a recent 
detection; Chadwick et al. 1998), all HBL. Celotti et al. (1998) and Protheroe
et al. (1997) (see also Catanese et al., these proceedings) estimate $\delta
\ga 10$ from MKN 421 and MKN 501, based on their TeV variability. It then
follows that for these sources $\theta \la 6^{\circ}$, smaller than average
value $\sim 20^{\circ}$ expected from the ``different viewing angle''
hypothesis (if HBL are uniformly distributed between $0^{\circ}$ and
$30^{\circ}$). Before drawing more general conclusions, however, we need to
have a larger sample of GeV/TeV HBL. The currently detected HBL, in fact,
could be $\gamma$-ray emitters precisely because they have larger than average
Doppler factors (and therefore smaller then average angles to the line of
sight).

It now seems accepted by most BL Lac researchers that orientation cannot be
the whole story in explaining the different properties of HBL and LBL.
Variations and expansions on the idea that the frequency of peak emission is
important have been presented at this meeting. 

Ghisellini and Fossati et al. (see also Fossati et al. 1997, 1998 and
Ghisellini et al. 1998) have suggested a scenario where $\nu_{\rm peak}$
anti-correlates with total power for all blazars. Namely, the synchrotron peak
of powerful sources (flat-spectrum radio quasars) is in the mm/far-IR band
while weaker sources (HBL) peak in the UV/X-ray band; LBL are in between. The
peak of the emission is related to electron energy, as $\nu_{\rm peak} \propto
B \gamma_{\rm break,e}^2$, with $\gamma_{\rm break,e}$ a characteristic
electron energy which is determined by a competition between acceleration and
cooling processes. Therefore, less powerful sources (where the energy
densities are relatively small) reach a balance between cooling and
acceleration at larger $\nu_{\rm peak}$, while in more powerful sources there
is more cooling and the balance is reached at smaller $\nu_{\rm peak}$.
Georganopoulos \& Marscher (these proceedings; see also Georganopoulos \&
Marscher 1998) also focus on $\nu_{\rm peak}$ but attribute a prominent role
to viewing angle and the electron kinetic luminosity, the latter being
(mildly) inversely dependent on $\nu_{\rm peak}$. As noted by Georganopoulos
\& Marscher, an anti-correlation between $\nu_{\rm peak}$ and power might 
help explaining the negative evolution of HBL. The basic concept is that if
higher powers are more common at higher $z$ (which is the general trend for
AGN and galaxies), then sources with higher $\nu_{\rm peak}$ (i.e., HBL) will
be discriminated against and will be more common locally.

I want to stress that the issues of the relation between HBL and LBL and of
which of the two classes is more numerous are very relevant to our
understanding of the physics of these objects. As mentioned above, for
example, the change of perspective from XBL/RBL to HBL/LBL has had important
implications for the accelerating jet model and has spurred a strong interest
in the study of the physical parameters underlying the emission processes in
BL Lacs. The deeper X-ray/radio surveys discussed at this meeting will be
vital to determine if nature favors high-energy peaked or low-energy peaked
BL Lacs. 

\begin{figure}
\psfig{file=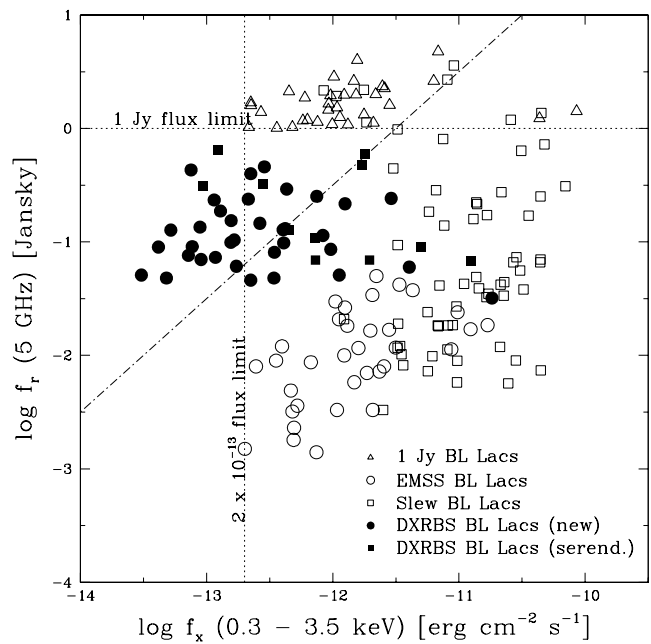,height=9truecm}
\caption[h]{The radio/X-ray flux plane for various samples of BL Lacs:
1 Jy (triangles), EMSS (open circles), Slew (open squares). Filled points
correspond to BL Lacs in the DXRBS survey (Perlman et al. 1998), both newly
discovered (filled circles) and serendipitous (filled squares). The dot-dashed
line corresponds to $f_{\rm x}/f_{\rm r} = 10^{-11.5}$ erg cm$^{-2}$ s$^{-1}$
Jy$^{-1}$. Note how DXRBS BL Lacs fill the gap between ``classical'' BL Lac
samples (and reach fainter X-ray fluxes).}
\end{figure}

If there is, as I (and many others) believe, a single population of BL Lacs
with a continuous range of peak frequencies, do we need to have the HBL/LBL
distinction? In reality, we do not. This was necessary when one was dealing
with the 1 Jy and EMSS samples, which we now know most likely represent the
extreme ends of a single population. BL Lacs with intermediate properties, in
terms of peak frequencies and $f_{\rm x}/f_{\rm r}$ ratios, have now been
discovered, as shown in Figure 1 (see also Laurent-Muehleisen et al. and
Wolter et al., these proceedings, and Perlman et al. 1998). The HBL/LBL
division, though, has probably a very interesting physical interpretation.
Padovani \& Giommi (1996) and Lamer, Brunner \& Staubert (1996) have studied
the {\it ROSAT} (0.1 -- 2.4 keV) spectra of BL Lacs and independently
discovered a dependence of the X-ray spectral index $\alpha_{\rm x}$ on
$\nu_{\rm peak}$ (or $f_{\rm x}/f_{\rm r}$) (see Figs. 2 and 6 of Padovani \&
Giommi 1996).  HBL, in fact, have on average steeper $\alpha_{\rm x}$ ($\sim
1.5$), which flattens for higher $\nu_{\rm peak}$, while the overall
broad-band spectrum is convex. LBL, on the other hand, have flatter
$\alpha_{\rm x}$ ($\sim 1.1$) and concave optical--X-ray continuum. This
points to different mechanisms being responsible for the (soft) X-ray emission
in the two classes, namely synchrotron and inverse Compton for HBL and LBL,
respectively. These X-ray studies suggest a dividing line between the two
classes at $f_{\rm x}/f_{\rm r} \sim 10^{-11.5}$ ($\alpha_{\rm rx}
\sim 0.8$ between 5 GHz and 1 keV) or $\nu_{\rm peak} \sim 10^{15}$
Hz. This picture is now being confirmed by {\it BeppoSAX} observations of BL
Lacs over a larger energy range (0.1 -- 10 keV; Wolter et al. 1998;
Padovani et al., these proceedings).

\section{Summary}
The main conclusions are the following: 

\begin{enumerate}

\item Unified schemes for BL Lacs are alive and well. There certainly are some
complications and problems with the basic scheme (which posits that all BL
Lacs are beamed FR Is) but these are only perturbations which cannot
invalidate the whole picture. 

\item The XBL/RBL (X-ray/radio selected) distinction is out-of-date and 
unphysical. BL Lacs most likely form one class, with a continuous distribution
of synchrotron peak energies, of which HBL (high-energy peaked BL Lacs, with
$\nu_{\rm peak} \ga 10^{15}$ Hz) and LBL (low-energy peaked BL Lacs) represent
the extreme ends. Orientation cannot play the major role in the differences
between the two extremes. 

\item There is a physical reason to adopt an HBL/LBL distinction: HBL are 
probably dominated by synchrotron emission in the soft X-ray band, while LBL 
have an inverse Compton component as well. 

\end{enumerate}

\acknowledgements

Most of my work on BL Lacs, on which the ideas expressed in this paper are
based, has been done in collaboration with Meg Urry and Paolo Giommi. Meg,
Paolo, Annalisa (Celotti), and Eric (Perlman), kindly read this paper at
various stages and provided very useful comments.

\end{document}